# In a hidden variables model all quantum observables must commute simultaneously.

James D. Malley[1]

[1]*Center for Information Technology, National Institutes of Health, Bethesda MD 20892*
*jmalley@helix.nih.gov*

Under a standard set of assumptions for a hidden variables model for quantum events we show that all observables must commute simultaneously. This seems to be an ultimate statement about the inapplicability of the usual hidden variables model for quantum events. And, despite Bell's complaint that a key condition of von Neumann's was quite unrealistic, we show that these conditions, under which von Neumann [1932] produced the first no-go proof, are entirely equivalent to those introduced by Bell [1966], Kochen and Specker [1967]. As these conditions are also equivalent to those under which the Bell-Clauser-Horne inequalities are derived, we see that the experimental violations of the inequalities demonstrate only that quantum observables do not commute. This same conclusion applies to the collection of elegant no-inequalities proofs devised by Greenberger-Horne-Zeilinger, Mermin, and Peres. Finally, we briefly consider alternative hidden variables models and how the problem of nonlocality is only imprecisely addressed by the usual model assumptions. Otherwise expressed, the usual hidden variables models have assumptions which collectively are too strong, and some of which, therefore, need modification or deletion.

PACS number(s): 03.65.Bz, 03.65.Ca

## I. INTRODUCTION

A wide range of no-go proofs for hidden variables (h.v.) models for quantum events have been developed and discussed over many years; see [1-8]. Some of these are related to Bell's inequalities, for which experimental demonstrations have been provided (and discussed at length). Others are of the so-called inequality-free type, and not requiring experiment; Mermin (1993) is an excellent resource for discussion of this type of proof.

Here we derive an alternative no-go proof with a rather striking and informative outcome: under the usual assumptions of an h.v. model *every* pair of observables must commute. As the conditions for an h.v. model studied here are known to be entirely equivalent to the conditions under which the usual Bell-Clauser-Horne (BCH) inequalities apply, we see that that the numerous experimental violations of the inequalities show only that quantum observables don't commute. The derivation shows, in effect, that study of the possible experimental loopholes of the Bell inequalities is not especially productive: only commutativity is on experimental test via the inequalities, so that even a perfectly loophole-free experiment has evidently little to justify its performance. Moreover, our derivation applies as well to the inequality-free type of no-go proofs: these elegant arguments show, again, only that quantum observables don't commute.

Our paper is organized as follows. We first consider specifications for a deterministic (or, factorizable stochastic) hidden variables models, such as are presented in [1, 2]. By extension of a result appearing in [14] we obtain our result on simultaneous commutativity. We conclude with a short discussion about hybrid h.v. models, such as those of [17, 18], that offer an alternative to the h.v. assumptions made here. These suggest a more promising route, should one be sought, for characterizations of quantum events as classical statistical schemes.

We include a rather lengthy appendix, dealing with details of the proof; this was not part of the original *PRA* paper.

We briefly anticipate some of this concluding discussion here. Thus, one of the more interesting consequences of our results is that the original conditions under which von Neumann [16] derived the first no-go proof for h.v. models are entirely equivalent to those introduced much later by Bell [7], Kochen and Specker [8]. Bell had criticized von Neumann for requiring the deterministic value assignment, under an h.v. model, to apply to sums over noncommuting (incompatible) observables, as well as for commuting ones; see [6; 12]. As an assignment for values across incompatible experiments seemed, to Bell, to be physically quite unrealistic, he introduced the less restrictive condition that the value assignment need apply only to across sums of commuting observables. However, we show that Bell's conditions (and those of Kochen-Specker) are in fact entirely equivalent to those of von Neumann, since, under an h.v. model of the Bell, or Kochen-Specker type, all observables must commute simultaneously.

## II. HIDDEN VARIABLES MODELS

Detailed specifications for a hidden variable (h.v.) model are given in [1, 2, 14], some of which we now recall. Let $Q = Q(H, D, \mathcal{O})$ denote a quantum system with Hilbert space $H$, quantum density operator $D$, and a family of observables $\mathcal{O}$.

Let $\Lambda = \Lambda(\Omega, \Sigma(\Omega), \mu)$ denote a classical probability space, where $\Omega$ is a nonempty set, $\Sigma(\Omega)$ is a Boolean $\sigma$-algebra of subsets of $\Omega$, and $\mu$ is a probability measure on $\Sigma(\Omega)$.

As used in this paper, a hidden variable model for a quantum system in a given state $D$ may make one or more of the following assumptions:

HV(a): Given $\omega \in \Omega$, $A \in \mathcal{O}$, there is a mapping $f$ from the pair $(\omega, A)$ to $\mathbb{R}$; it is required that the value of $f(\omega, A)$ be an eigenvalue of $A$;

HV(b): For any two commuting observables $A$, $B$, the mapping $f$ is such that

$$f(\omega, A+B) = f(\omega, A) + f(\omega, B). \tag{2.1}$$

HV(c): The measure $\mu$ correctly returns the marginal probabilities for each observable $A$, that is for any real Borel set $S$, $\mu$ is such that

$$tr[DP_A(S)] = \int_\Omega f(\omega, P_A(S))\, d\mu, \tag{2.2}$$

where $P_A(S)$ is the projector associated with set $S$ in the spectral resolution for $A$.

HV(d): For any two commuting observables $A$, $B$, the measure $\mu$ correctly returns the joint probabilities; that is, for $S$, $T$ real Borel sets, the measure $\mu$ is such that

$$tr[DP_A(S)P_B(T)] = \int f(\omega, P_A(S)P_B(T))\,d\mu, \tag{2.3}$$

for $P_A(S), P_B(T)$ the projectors associated with sets *S, T* in the spectral resolutions of *A, B*, respectively.

Next we recall a discussion and a definition from [14] on classical and quantum conditional probability. Assume there is a classical probability space such that outcomes for projectors *A, B* can be described by a joint distribution $\mu$. It is interest to ask when the conditional distribution derived from $\mu$ agrees with the standard definition of quantum conditional probability; see [13] and the *Appendix* given below for details of the probability background. For projectors *A, B*, and any quantum state *D*, the quantum conditional probability of *A*, given *B*, is defined by:

$$\Pr[A \mid B] \;=\; tr[DBAB]/tr[DB]\,. \tag{2.4}$$

Consider now the two conditional distributions, that derived from $\mu$ and that derived from the standard definition (2.4) above. When these are equal we will say that the *conditional probability rule* holds. For any projector *X*, let

$$X^{-1}(1) = \{\omega \in \Omega : X(\omega) = 1\}. \tag{2.5}$$

Then as shown in [14; Theorem 1]:

*Theorem 1.* Assume *dim H* ≥ 3, and that HV(a), HV(c), HV(d) hold. Then for one-dimensional projectors *A, B,* the conditional probability rule holds:

$$\mu[a \mid b] = \mu[a \cap b]/\mu[b] = tr[DBAB]/tr[DB], \tag{2.6}$$

where $a = A^{-1}(1), b = B^{-1}(1)$.

We re-state the proof given in [14] in the *Appendix*.

We observe that the restriction of this result, to one-dimensional projectors, is not required but the proof in this case can be obtained using straightforward inner product vector space methods; see [14] and Gudder [15; Corollary 5.17]. We do not argue here that the no-go proof presented below, based on this restricted case, is in any sense technically simpler than the original Kochen-Specker or Bell proofs---this is partly a matter of taste. However, we will argue that the end-point of the proof presented here, namely commutativity, is more informative and transparent as regarding the problems with local h.v. models, in particular those studied using the BCH inequalities.

We also note that, as discussed in [14], there are two other conditions equivalent to HV(b): a *Borel function rule*, and a *product rule*, both introduced in [2]. Any of these three choices will suit the purposes of our discussion.

In [1] the set of conditions HV(a), HV(c), HV(d) is called a *deterministic hidden variables model* (equivalently, a *factorizable stochastic model*). To be more precise, in this paper we take the three conditions {HV(a), HV(c), HV(d)} to jointly define an *h.v. model.* As shown in [1] the conditions {HV(a), HV(c), HV(d)} are also entirely equivalent to {HV(a), HV(b), HV(d)}, and these are the conditions introduced by Bell [7], Kochen-Specker [8]. Moreover, as shown by Fine [3; Proposition (2)], a necessary and sufficient condition for the existence of a deterministic h.v. model is that the usual BCH inequalities must hold. Van Fraassen [10; 102-105] gives further details of the Fine results, showing how locality, in the form of factorizability, is built into Fine's definition of h.v. models. Further details concerning how locality might be differently defined can be found in Fine [18; Appendix to Chapter 4].

## III. HIDDEN VARIABLES AND COMMUTATIVITY

*Theorem 2.* Assume $\dim H \geq 3$ and that an h.v. model holds for quantum events. Then all quantum observables commute.

*Proof.* Let *A, B* be two quantum observables. Without loss of generality we may assume they are one-dimensional projectors: *A, B* commute if and only if all projectors appearing in their spectral resolutions commute, and all the projectors may be re-expressed as (non-unique) sums of one-dimensional ones. From *Theorem 1* we have that

$$\mu[a,b] = \mu[a \mid b] \cdot \mu[b] = \{tr[DBAB]/tr[DB]\} \cdot tr[DB] = tr[DBAB], \tag{3.1}$$

and also that

$$\mu[a,b] = \mu[b \mid a] \cdot \mu[a] = \{tr[DABA]/tr[DA]\} \cdot tr[DA] = tr[DABA]. \tag{3.2}$$

Hence

$$tr[DBAB] = tr[DABA] \tag{3.3}$$

for all density operators *D*. Thus *BAB* = *ABA*. From this, and using $A^2 = A, B^2 = B$, we easily show that

$$(AB - BA)^2 = 0. \tag{3.4}$$

Since $C = AB - BA$ is skew-Hermitian, $C^2 = 0$ implies $C = 0$, and the result is proven.

An alternative proof appears in the *Appendix*, where we also discuss the key technical ideas of the proof.

We note that Fine [3; Theorem 7] obtained a commutativity result using a rather different condition, called the *joint distribution (jd) condition*. Briefly, this states that a measure space be given which returns the correct marginal distributions for a set of (not necessarily commuting) observables, $A_1, A_2, ..., A_k$, and which also reproduces the marginal for any observable of the form $f(A_1, A_2, ..., A_k)$, for any Borel measurable *f*. The joint distribution condition does not by itself reference h.v. models, but might be considered as useful background to the problems with such models. More precisely, the h.v. conditions given above, HV(a), HV(c), HV(d), do not in any obvious way validate the Borel function requirement, just stated, in the *jd condition*. On the other hand, we have from above that an h.v. model is equivalent to simultaneous commutativity for all observables, so the *jd condition* is now seen as an interesting alternative for the collected assumptions of a deterministic h.v. model.

## IV. DISCUSSION

We have shown that under the standard h.v. model assumptions ($dim\ H \geq 3$), all quantum observables must commute. Seemingly no more sharply informative no-go proof is possible, and the conclusion obtains under the Bell, Kochen-Specker, or Fine conditions for an h.v. model. In particular, we see that the sum rule HV(b) is valid for noncommuting observables, in the presence of the other conditions for an h.v. model, namely HV(a), and HV(d). The requirement that HV(b) apply for noncommuting observables was made by von Neumann in his original 1932 no-go proof for h.v. models. This was declared by Bell to be entirely unphysical for any plausible h.v. model for quantum events, and he preferred to assume HV(b) only for commuting observables; see the discussion in [6, 12]. In fact, we now see that von Neumann's h.v. assumptions were no more or less unphysical than were Bell's, or Kochen-Specker's apparently less restrictive set of assumptions. In this sense von Neumann's original proof is vindicated.

Finally, given the above it seems appropriate to urge consideration instead of models for quantum events that are not tied to these h.v. conditions. Effectively, the Bell, Kochen-Specker, and the (now equivalent) von Neumann, conditions, are still too restrictive and truly weaker models could to be considered. Such hybrid models appear to be already at hand, as in [17]; see also the discussion of *prism models* in [18] and the references to the literature cited therein. A significant change presented by these prism models is that the hidden variables are not assumed to be factorizable, but do satisfy what Fine calls

*Bell-locality*, an assumption briefly described as "no outcome-fixing action-at-a-distance"; see [18; Appendix to Chapter 4]. Under this construal, violations of the BCH inequalities do not constitute a failure of Bell-locality, and our no-go commutativity result does not extend to a negation of Bell-locality. Using the simplified notation of Fine, Mermin and others, a central question then is, with what do we replace the product rule, $(AB)(\square) = A(\square) \cdot B(\square)$, in order to capture a notion of locality?

**Appendix: Technical details and a discussion of the proofs of Theorems 1 and 2.**

We require the following definitions:

*Definition 1*. For two projectors *A*,*B*, we write $A \le B$ to mean that $AB = BA = A$.

*Definition 2*. The h.v. product rule: For any two commuting observables *A*, *B*, the mapping $f$ (in HV(a) above) is such that $f(\lambda, AB) = f(\lambda, A) \cdot f(\lambda, B)$.

As noted in the text above, the product rule is equivalent to the sum rule HV(b), in the presence of HV(a) and HV(c). A beautiful and useful network of other equivalences is given in Fine [1]. We note that Fine and Mermin use an alternative notation for the mapping $f$ which we now adopt (and probably should have much earlier, starting in [14]): for projector observable *A* we write: $f(\lambda, A) = A(\lambda)$. The product rule, for example, then has the smoother statement:

$$(AB)(\lambda) = A(\lambda)B(\lambda)$$

for commuting projectors *A, B*.

We now prove the key lemma needed for *Theorem 1*. Recall the notation used in the text: given an h.v. model, and any projector *X*, let

$$x = X^{-1}(1) = \{\lambda \in \Lambda : X(\lambda) = 1\}.$$

*Lemma 1*. In an h.v. model, if projectors *A, B* are such that $A \le B$ then $a \cap b = a$.

*Proof*. From the product rule we have $(AB)(\lambda) = A(\lambda)B(\lambda)$. If it is true that $A(\lambda) = 1$, then using the spectrum rule HV(a), it is also true that $B(\lambda) = 1$, so that

$$a \cap b = \{\lambda \mid A(\lambda) = 1 \text{ and } B(\lambda) = 1\} \supseteq \{\lambda \mid A(\lambda) = 1\} = a.$$

Since it is always the case that $a \cap b \subseteq a$ the result follows.

Apart from the existence of a basic phase space and the equality of marginals

(= HV(c)), *Lemma 1* is *only* place where the truly key element---the product rule---of an h.v. model is invoked. As noted in Fine [18] this factorizability is often considered to be the defining condition for quantum locality. However, from *Theorem 2*, we will argue that it is simply too strong an assumption: only a trivial conclusion---commutativity---follows from the product rule.

Let's consider how a classical probability space (*phase space*) might further relate to quantum events. If only marginal distributions are of interest then a phase space is, trivially, available for all the quantum observables, free of any h.v. conditions, and we note that on any such space the joint probability for any pair of variables always exists, as the simple product measure. Also, from the h.v. assumptions we know that the marginals derived from the phase space must agree with those specified for the quantum system. However, we don't, as yet, know how the joint on the phase space relates (if at all) to any distributions specified by the quantum system. To make this connection we require a result that appears as an *Exercise* in [13]. As this masterful text is apparently out of print, we provide it here. To make the discussion more complete, we require some definitions and a result from matrix algebra. First, the projector lattice definitions:

For Hilbert space *H,* let *P(H)* denote the collection of projectors. For any projector *A* we write $A^{\perp} = I - A$ to denote the *orthocomplement* of *A*; projectors *A, B* are *orthogonal* if $AB = BA = 0$. A collection of projectors is called *orthogonal* if all pairs in the collection are orthogonal. For a countable collection of orthogonal projectors $\{A_i\}$ we write $\oplus A_i$ for their sum. Although we don't specifically need the context, the set of all projectors *P(H)* forms a lattice with numerous, nice features: it is orthocomplemented, ortho-modular, separable, has the covering property, and is both atomic, and atomistic; see Chapter 10 in [13]. We do need the following definition:

*Definition 3*. Given the collection (a *lattice*) of projectors *P(H)*, we say that a real-valued function $\mu$ is a probability measure on *P(H)* if

*(i)* $0 \le \mu(A) \le 1$, for all $A \in P(H)$, $\mu(0) = 0$, $\mu(I) = 1$;

*(ii)* $\mu$ is countably additive; for every countable orthogonal sequence $\{A_i\}$ the series $\sum \mu(A_i)$ converges and $\mu(\oplus A_i) = \sum \mu(A_i)$.

We state the fundamental result of Gleason (1957); see [13]. We will subsequently work to show the uniqueness of the density operator:

*Gleason's Theorem*. If dim$H \geq 3$ then every probability measure on *P(H)* arises from a density operator *D* on *H*, and is of the form $\Pr[B] = tr[DB]$.

Next, a result from matrix algebra; for a real or complex, square matrix *A* we write $A^*$ to denote its conjugate-transpose.

*Definition 4*. A matrix *A* is said to be *normal* if $AA^* = A^*A$

The main examples of normal matrices are those *A* that are *hermitian*, $A = A^*$, or *skew-hermitian*, $A^* = -A$. A matrix *A* is *unitary* if $AA^* = A^*A = I$. The following is a classical result, due to Shur (1909) and Toeplitz (1918), the *spectral decomposition* of normal operators [see for example [19], p. 174-180]:

*Lemma 2*. A square matrix is normal if and only if it is unitarily similar to a diagonal matrix of its eigenvalues: $A = UDU^*$, where $D = diag[\lambda_i]$, with $\{\lambda_i\}$ the eigenvalues of $A$, where *U* is unitary, and the columns of *U* may taken to be the eigenvectors of *A*.

The next result will be used repeatedly. Recall that a matrix $D$ is *positive* if $\langle h, Dh \rangle \geq 0$ for all vectors $h \in H$. By construction, the density operator $D$ for a quantum system is assumed to be positive. Then:

*Lemma 3*. Given a normal operator $B$ such that $tr[DB] = 0$ for all positive operators $D$, it follows that $B = 0$.

*Proof*. Use the spectral decomposition of $B$ and let $D$ range over all one-dimensional projectors corresponding to eigenvalues of $B$.

We can now begin the proof of the key *Exercise*:

*Exercise. [13; p. 288]*. For Hilbert space $H$, let $\mu$ be a probability measure on the lattice of projectors $P(H)$, and let $B$ be any projector such that $\mu(B) \neq 0$. Then there exists a unique probability measure on $P(H)$, which we denote by $\Pr_\mu(\cdot \mid B)$, such that for all projectors $C \leq B$ it is the case that $\Pr_\mu(C \mid B) = \mu(C)/\mu(B)$.

*Proof*. By Gleason's theorem we know at once that, for any projector $A$, $\Pr_\mu(\cdot \mid B)$ must be of the form $\Pr_\mu(A \mid B) = tr[D_\mu A]$ for some density operator $D_\mu$. If we can show that $D_\mu = D$ (= the ambient density for our quantum system) then the proof would be finished. Therefore, suppose now that $D_1$ and $D_2$ are two density operators such that

$$tr[D_1 C] = tr[D_2 C] = \mu(C)/\mu(B)$$

for all projectors $C$ such that $C \leq B$.

We have that

$$tr[D_1 B] = tr[D_2 B] = 1$$

and

$$tr[D_1 B^\perp] = tr[D_2 B^\perp] = 0$$

for $B^{\perp}$ the orthogonal complement of $B$, $B^{\perp} = I - B$. Consider now any unit vector $\square$ in the range of $B^{\perp}$: $\square$ is a 1-eigenvector of $B^{\perp}$. As $B^{\perp}$ is a projector we may write it as a sum of one-dimensional projectors, one of which is the projector defined by $\square$, $P_{\square} = |\square\rangle\langle\square|$. Hence

$$B^{\perp} = P_{\square} + P_{\square} + \ldots + P_{\square}$$

where $\square, \square, \ldots, \square$ are all unit vectors. For any density operator $D$ it follows that

$$0 = tr[DB^{\perp}] = tr[DP_{\square}] + tr[DP_{\square}] + \ldots + tr[DP_{\square}].$$

As $D$ is positive (by definition) we get $tr[D|\square\rangle\langle\square|] = \square^{*} D\square \geq 0$, all vectors $\square$, and it follows that each term in this sum must be zero. In particular, we have $D\square = 0$. Hence in fact $D_1\square = D_2\square = 0$ for all vectors $\square$ in the range of the projector $B^{\perp}$. Writing

$$\square = \{v \mid v = Bh, h \in H\}, \text{ and } \square^{\perp} = \{\square \mid \square = B^{\perp}g, g \in H\}$$

we see that

$$H = \square \oplus \square^{\perp}.$$

Next, let $R$ be any one-dimensional projector in *P(H),* so that $R = |\square\rangle\langle\square|$ for some unit vector $\square$. Then

$$tr[D_1R] = \langle\square \mid D_1\square\rangle, \text{ and } tr[D_2R] = \langle\square \mid D_2\square\rangle.$$

We can always decompose $\square$ as

$$\square = \square_1 + \square_2 \text{ where } \square_1 \in \square, \ \square_2 \in \square^{\perp}.$$

Consequently

$$\begin{aligned}\langle\square, D_i\square\rangle &= \langle\square_1, D_i\square_1\rangle + \langle\square_2, D_i\square_2\rangle + \langle\square_1, D_i\square_2\rangle + \langle\square_1, D_i\square_2\rangle \\ &= \langle\square_1, D_i\square_1\rangle + \langle\square_2, D_i\square_2\rangle + 2\mathrm{Re}\langle\square_1, D_i\square_2\rangle \quad \text{for } i = 1,2.\end{aligned}$$

We also have $D_i\square_2 = 0$. Hence

$$tr[D_i R] = \langle \psi_1, D_i \psi_1 \rangle = \|\psi_1\|^2 tr[D_i R_1],$$

where $R_1$ is the projector onto the one-dimensional subspace generated by $\psi_1$. Note that $R_1 \le B$, so that by assumption

$$tr[D_1 R_1] = tr[D_2 R_1].$$

From this it now follows that

$$tr[D_1 R] = tr[D_2 R]$$

for all one-dimensional projectors *R*. Consider, finally, the skew-hermitian operator $G = D_1 - D_2$. As *G* is a normal operator, and a one-dimensional projector *R* is always positive, we use *Lemma 3* to conclude that $G = 0$, so $D_1 = D_2$. This finishes the proof of the *Exercise*.

The central idea of the proof above is the uniqueness feature of Gleason's theorem, a fact apparently not widely appreciated or utilized. Let's also clarify the appearance of the projector pairs *A, B* in the result. Thus, if *D* is the ambient density operator for the quantum system, then Gleason's theorem provides for *some* density $D_\mu$ that correctly matches the conditional probability $\Pr_\mu(\cdot \mid B)$ defined on those random variables in the phase space that correspond to projectors. If this probability is also such that whenever $C \le B$ we also must have $\Pr_\mu(C \mid B) = \mu(C)/\mu(B)$, then the density is unique,

$$D_\mu = D_B = BDB / tr[DB]$$

and

$$\Pr_\mu(A \mid B) = tr[D_\mu A] = tr[D_B A], \text{ for all projectors } A.$$

The important point here is that uniqueness does not apply to just projectors *C* such that $C \le B$: uniqueness of the probability measure for *all* projectors *A* obtains for $\Pr_\mu(\cdot \mid B)$, if for every C that $C \le B$, then $\Pr_\mu(C \mid B) = \mu(C)/\mu(B)$. Of course, the uniqueness applies as well to the unconditional probably measure $\Pr(A) = tr[DA]$, defined for all projectors $A$, since trivially, $AI = IA = A$, for identity operator $I$.

We now have the materials to prove our first main result:

*Theorem 1.* Assume dim $H \geq 3$. Then in a deterministic h.v. model the conditional probability rule holds: $\square[a \mid b] = \square[a \square\, b]/\square[b] = tr[DBAB]/tr[DB]$.

*Proof.* To this end, consider two projectors $C$, $B$ such that $C \square B$. We have

$$tr[DBCB]/tr[DB] = tr[DCB]/tr[DB] = tr[DC]/tr[DB],$$

or, re-stated:

$$\Pr_D[C \mid B] = \Pr_D[C]/\Pr_D[B].$$

By the Lemma above we also have

$$\square[c \mid b] = \square[c \square\, b]/\square[b] = \square[c]/\square[b].$$

From the h.v. rules we have equality of the marginals:

$$\square[c] = tr[DC] = \Pr_D[C], \text{ and } \square[b] = tr[DB] = \Pr_D[B].$$

Hence

$$\square[c \mid b] = \Pr_D[C \mid B]$$

for all projectors such that $C \square B$. The proof will be complete if we can show that the agreement of the phase space conditional probability with the usual quantum system trace-rule probability implies agreement for all projector pairs, $C \square B$. However, this follows at once from the *Exercise* above.

Finally, we provide an alternative proof for *Theorem 2*. We begin with the observation that Gleason's theorem does not require the projectors in question to be one-dimensional. We made this restrictive assumption, that they were so, in our earlier paper [14], as a means to simplify the proof of *Theorem 1* appearing there. However, we now don't think that any useful simplification results: using Gudder's vector space argument is instructive but isn't a significant requirement relative to our goals here. Instead we can use a method relating to an early result of Davies [20] (more on this below). Thus:

*Theorem 2 (Alternative Proof).* Assume $\dim H \geq 3$ and that an h.v. model holds for quantum events. Then all quantum observables commute.

*Proof.* Assume an h.v. model holds, and let $A$, $B$ be any two projectors. As in the first half of the original proof of *Theorem 2* above, we find that $ABA = BAB$. As the projectors need not be one-dimensional, this conclusion also holds for any two projectors from the set $\{A, B, I - A, I - B\}$. Now, write

$$\tilde{A} = I - A, \tilde{B} = I - B.$$

Then

$$A = A(B + \tilde{B})A = ABA + A\tilde{B}A = BAB + \tilde{B}A\tilde{B},$$

since $A^2 = A, B^2 = B,$ and $A\tilde{A} = \tilde{A}A = B\tilde{B} = \tilde{B}B = 0.$

Thus

$$AB = BAB^2 + \tilde{B}A\tilde{B}B = BAB,$$

as well as

$BA = B^2AB + B\tilde{B}A\tilde{B} = BAB.$ Thus $AB = BA$ and the proof is complete.

Let's comment now on the result of Davies [20]. Consider a product space of events of the form $B$ and $A \mid B$, where we connect events (real Borel sets) and projectors in the usual way. We can use the ambient probability given by the density $D$ and the conditional probability $\Pr[A \mid B]$ given as usual by

$$\Pr[A \mid B] = tr[D_B A]$$

to define a joint product probability on the pair $\{A;B\}$. This is a valid classical statistics construction, and nothing new that is specifically quantum is being invoked here. Davies used this construction to discuss *instruments* and *effects*, and using a proof very similar to the second half of the alternative proof of *Theorem 2* above, he showed that

$$\Pr\{A;B\} = \Pr\{B;A\} \text{ if and only if } AB = BA;$$

see Gudder [15; p. 87]. This technical similarity of the second half of our *Theorem 2* and the early result of Davies is entirely classical: it does not require Gleason's theorem nor have implications for h.v. models.